# The Power Spectrum of Microwave Background Temperature Anisotropies Measured by the Tenerife Experiment.

F. Atrio–Barandela

*Física Teórica. Facultad de Ciencias.*
*Universidad de Salamanca, 37008 Spain.*
*e–mail: atrio@astro.usal.es*

S. Gottlöber and J.P. Mücket

*Astrophysikalisches Institut Potsdam.*
*An der Sternwarte 16. D-14482 Potsdam, Germany.*
*e–mail: (sgottloeber,jpmuecket)@aip.de*

## ABSTRACT

We determine the slope of the power spectrum of the matter density perturbations from the Tenerife observations of the cosmic background radiation temperature anisotropies. We compute the projected radiation anisotropy power spectrum measured by this experiment and study its dependence with respect to the slope of the temperature anisotropy power spectrum on the sky. We show that Tenerife alone implies the upper bound on the spectral index of $m \leq 3$. Stronger conclusions can not be reached due to the small data set. The method proposed can be applied to any small scale experiment. Sampling the same region of the sky with different window functions could probe the slope of the radiation anisotropy power spectrum at different scales and confirm the presence of Doppler peaks.

*Subject headings:* Cosmic Microwave Background. Cosmology: theory. Methods: numerical

## 1. Introduction.

In the last few years several groups have reported measurements of temperature anisotropies on the Cosmic Microwave Background (CBR) at different angular scales (see White, Scott & Silk 1994; Bond 1997, for a review). These measurements are becoming sensitive enough to disproof models of large scale structure and to measure cosmological parameters such as the amplitude and slope of matter density perturbations (Bennett et al. 1996), the height and location of the first Doppler peak in the temperature anisotropy power spectrum (Scott et al. 1995), and to determine the possible existence of a cosmological constant (Lineweaver et al. 1996). Assuming temperature fluctuations of the CBR to be Gaussian distributed, the power spectrum of the temperature anisotropies contains all the physical information about the temperature field. The power spectrum has been obtained for the COBE/DMR full sky temperature anisotropy map, where window function facilitates the decomposition of the temperature field in its spherical coefficients (Bunn, Hoffman & Silk 1996). The variance of data from small scale observations has been used to constrain models of Large Scale Structure (Ganga et al. 1996; Gorski et al. 1996; White et al. 1996).

To the best of our knowledge, no attempt has been made to estimate the slope of the radiation anisotropy power spectra for experiments with partial sky coverage but higher resolution than COBE/DMR. In this article we shall perform such analysis aiming to probe the overall shape of the power spectrum at intermediate and small angular scales. We shall center our discussion on the analysis of one–dimensional data sets and, in particular, on the scan measured by the Tenerife group (Hancock et al. 1994). The situation is common in cosmology where information about the matter power spectrum is inferred from one–dimensional pencil beam galaxy surveys (Broadhurst et al. 1990) or from the angular correlation function of galaxies (Maddox et al. 1990). The outline of the paper is as follows: in Section 2, we show that for small scale experiments the power contained in the data is related to the temperature anisotropy power spectrum on the celestial sphere. In Section 3 we will assume that the radiation anisotropy power spectrum is given by a power law with slope $m$, roughly constant on the scales probed by the Tenerife experiment. We shall show how this slope can be measured from the data.

In Section 4 we apply the technique devised in the preceeding section to the temperature anisotropies observed at Tenerife. We perform several Monte Carlo simulations to determine the confidence level of our estimates. Finally, in Section 5 we discuss our results and present our main conclusions.

## 2. Power Spectrum of one-dimensional temperature anisotropy fields.

If temperature anisotropies on the sky are Gaussian distributed, they are fully described by the angular correlation function

$$C(\theta) = \frac{1}{4\pi} \sum_l (2l+1) C_l W_l P_l(\cos\theta), \quad (1)$$

where $C_l = |a_{lm}|^2$ is the radiation anisotropy power spectrum and $W_l$ is the window function for a particular experiment. In the limit $l \gg 2$ the correlation function can be rewritten as (Bond & Efstathiou 1987):

$$C(\theta) = \frac{1}{2\pi} \int l \, dl \, C(l) W(l) J_o(l\theta). \quad (2)$$

$J_o$ denotes the cylindrical Bessel function, and $C(l), W(l)$ represent continuos functions that verify $C(l) = C_l, W(l) = W_l$ at positive integers. This limit is equivalent to consider the sky as a flat surface and within the same approximation the temperature anisotropy can be expressed as

$$\left(\frac{\Delta T}{T_o}\right)(\mathbf{r}) = \int \frac{d^2k}{(2\pi)^2} \frac{\delta T}{T_o}(\mathbf{k}) e^{-i\mathbf{k}\mathbf{r}}, \quad (3)$$

where $\mathbf{r}$ is a direction on the sky. The relation between the "continuos radiation anisotropy power spectrum" defined by $P_{rad}(k) = |\delta T/T_o(\mathbf{k})|^2$ and $C(l)$ is

$$k^2 P_{rad}(k) = l^2 C(l) W(l). \quad (4)$$

In this last expression, angular and linear scales are related by $r = R_H \theta, l = k R_H$, being $R_H$ the horizon radius that, for convenience, we shall take equal to unity.

Most observations on small angular scales sample a small fraction of the sky. In general, the experiments measure the temperature along a single angular direction. In the flat sky approximation the temperature anisotropy along, for example, the $x$ axis is

$$\left(\frac{\Delta T}{T_o}\right)(x) = \int \frac{d^2k}{(2\pi)^2} \frac{\delta T}{T_o}(\mathbf{k}) e^{-ik_1 x}. \quad (5)$$

All the statistical information contained in the data is determined by the one–dimensional power spectrum $P_{1D}(q) = |\delta T/T_o(q)|^2$, being $\delta T/T_o(q)$ the 1D Fourier transform of the data. Since the correlation function is common to both the 2D and 1D temperature field because of statistical isotropy, the projected power along one direction and the sky power spectrum are related by the integral equation (Peacock 1992)

$$P_{1D}(q) = \frac{1}{\pi} \int_o^\infty dk_2 P_{rad}(\sqrt{q^2 + k_2^2}). \qquad (6)$$

In our approximation, equation (6) can also be expressed as

$$P_{1D}(q) = \frac{1}{\pi} \sum_{l>q} C(l) W(l). \qquad (7)$$

The last identity follows from equation (4). An immediate consequence of this equation is that $P_{1D}(q)$ must be a decreasing function of wave number.

In principle, inversion of equation (7) would give the radiation anisotropy power spectrum on the celestial sphere $C(l)$. This type of inverse problems are common in astronomy and are usually ill-conditioned (Lucy 1994). The number of data points is usually insufficient to recover the power. Furthermore, observational errors in the data are amplified by any inversion procedure and the final answer is totally unreliable. Instead of trying to recover the full power spectrum, we will assume it to be described by a power law at the scales where the window function is non-negligible and we shall discuss in the next section how the slope can be recovered from the data.

## 3. The slope of radiation anisotropy power spectrum at Tenerife scales.

As mentioned in the introduction, we will center our study on the Tenerife experiment. For most cosmological models with $\Omega_o = 1$, temperature anisotropies at Tenerife scales are dominated by fluctuations in the gravitational potential (Sachs & Wolfe 1967). For matter spectral indexes $n \leq 3$ the relation is given by (Peebles 1982)

$$C_l = (a_2)^2 \frac{\Gamma[l + (n-1)/2]\Gamma[(9-n)/2]}{\Gamma[l + (5-n)/2]\Gamma[(3+n)/2]}. \qquad (8)$$

The quadrupole $a_2$ will be taken here as a normalization constant. The radiation anisotropy power spectrum in the region $2 << l \leq 50$ is essentially a power law: $P(k) \sim k^m$ and its slope $m$ is related to the matter spectral index $n$ by: $m \approx n - 3$.

One can estimate quantitatively the dependence of the radiation anisotropy power spectrum $P_{1D}(q)$ with $m$ on scales $20 < l \leq 40$. In that region, the power spectrum is $P_{rad}(k) \approx k^m \exp(-k^2 \sigma^2)$ because the window function is dominated by the exponential cut–off introduced by the finite beam size ($\sigma$) of the antenna. Within this approximation the projected 1D power spectrum defined by (eq. [6]) is

$$P_{1D}(q) = \frac{1}{\pi} e^{-q^2 \sigma^2} \int_o^\infty dy (y^2 + q^2)^{m/2} e^{-y^2 \sigma^2}. \qquad (9)$$

This integral can be calculated in closed form in terms of Whittaker functions (Gradshteyn & Ryzhik 1980):

$$P_{1D}(q) = \frac{\sigma^{-\frac{m+3}{2}}}{\sqrt{\pi}} e^{(-\frac{1}{2}\sigma^2 q^2)} q^{\frac{1}{2}(m-1)} W_{a,-a}(q^2 \sigma^2), \qquad (10)$$

with $a = (1+m)/4$. In the limit $q\sigma \geq 1$, the dependence of the power on $q$ reduces to:

$$P_{1D}(q) \propto q^m \exp(-\sigma^2 q^2). \qquad (11)$$

This is our main result of this section: at high multipoles the projected power spectrum $P_{1D}(q)$ depends on the spectral index in the same way as the radiation anisotropy power spectrum $P_{rad}(k)$. The 1D power is boosted by the projection of short wavelengths in two dimensions, like in pencil beam surveys (Kaiser & Peacock 1991), but the overall shape remains unchanged. This result should not come as a surprise since it is a consequence of the statistical isotropy of the correlation function. Nevertheless, it has an interesting application to observations. On the region dominated by the exponential tail of the window function the amount of information contained in a 2D map or in a scan is similar. Therefore, experiments like Tenerife or Bartol–IAC (Piccirillo et al. 1996) that obtain data in the form of a single scan can constrain the radiation anisotropy power spectrum at the same confidence level than maps do. The main limitation does not come from the geometry of the experiment, but by the number of independently measured data points.

Since we will analyze the data measured by the Tenerife group, we will restrict our analysis to this experiment. The Tenerife group is continuously measuring a strip of 360 degrees in R.A., at declination $40^\circ$. Their observational strategy is a double beam switch with an $8^o$ chop in right ascension with $5.5^o$ beam width. The window function of the experiment peaks at $l = 20$ and is rather insensitive to multipoles outside the range $10-50$. Their final data is displayed as a scan of fixed declination and right ascension from 12 to 18h. Outside this range, the signal is dominated by the galactic contribution and other systematic errors. For the Tenerife experiment we shall use information contained in the range $20 < l \leq 40$. For small multipoles the curvature of the sky renders our treatment of the data invalid. For multipoles larger than $l = 40$ the window function erases the cosmic signal and the data contains only uncorrelated noise. In the region of interest, $\log(P_{1D}(q))$ is almost a straight line. In Figure 1 we plot the projected power spectrum for different slopes: $m = -2.5, -2, -1.5, 1$. The corresponding spectral index of matter power spectrum is approximately $n \simeq m + 3$. The 1D power spectrum was computed using equations (7) and (8) for $m < 0$ and equation (6) for $m \geq 0$. The slope $m_{1D} = \Delta \log(P_{1D}(q))/\Delta q$ was obtained by means of a least square fit to the one dimensional power spectrum in the region $l = 20 - 40$. More than 50% of the total power measured by the experiment is contained in this region of $l$ space. In column 2 of Table 1 we quote the figures for different slopes. We verified that the numerical value and the analytical estimate using equation (11) were very similar.

It is not surprising to find that $m_{1D}$ depends weakly on $m$. At high multipoles, the power is dominated by the finite beam width and the signal is exponentially damped. Even though, we shall see in the next section that meaningful statistical information can be obtained from the data.

## 4. Estimation of the radiation spectral index from Tenerife data.

Our procedure to estimate the slope of the matter power spectrum measured by the Tenerife experiment is very simple. We compute the 1D power spectrum from the data by means of a Fourier transform. The slope $m_{1D}$ is obtained by a least square fit to $\log P_{1D}(q)$ in the region $20 < l < 40$. We compare this value with the theoretically expected value that can be read from Table 1.

To estimate the statistical uncertainty associated to the numbers given in Table 1, we perform Monte Carlo simulations by selecting five hundred Tenerife scans randomly placed on a single realization of the sky for any given $n = m + 3$. We computed the 1D power in order to estimate $m_{1D}$. Each scan consisted of 90 points separated by $1^o$ in R.A. This data was later binned to obtain an independent point per beam width. Our simulations were run considering zero noise and were repeated by adding noise levels of 20, 50 and 100% of the cosmic signal to each scan. The slopes were Gaussian distributed around the mean, given in Table 1. The first column gives the slope of the radiation anisotropy power spectrum. In the second and third columns we write the theoretical expected value of $m_{1D}$ and the value for our particular realization of the sky. On columns 4 to 7 we give the measured slope for different noise levels. One should notice that the effect of white noise in our simulations is to increase the slope (to decrease it in absolute value). This is easily understood since white noise has a slope $m = 0$ and with increasing ratio $\sigma_n/\sigma_s$ the scan has a larger contribution from larger slopes.

The shortcomings of power spectrum analysis based on periodograms or other types of spectrum estimator are well known (Press et al. 1992). For example, according to equation (7) $P_{1D}(q)$ must be a decreasing function of $q$ and in general it was not. Noise and data discretness and aliasing between different scales induce oscillations in the one dimensional power that results in a very high (of the order of 0.8) r.m.s. deviation ($\tilde{\sigma}$) on the values of columns 4–7 of Table 1. With this result, it would be impossible to distinguish between $m = -2$ and $m = 1$ models at $1\tilde{\sigma}$ confidence level, even for a noise free scan. A partial solution to the problem can be obtained by averaging power spectra from several data sets (see also Press et al. 1992). Therefore, we repeated our simulations but averaging the power of $N$ different scans before computing the slope. We took $N = 3, 5, 10$ and the result was that the mean slope of the 500 scans did not change and the variance was reduced by a factor proportional to $\sqrt{N}$. The values given in Table 1 were obtained after averaging three scans before computing the power. The r.m.s. deviation was $\tilde{\sigma} \approx 0.04$ and was never larger than 0.05 in all cases.

We finally apply our formalism to the data. The Tenerife group has published the data on three different frequencies at declination $40^o$ and soon will be releasing their measurements for two more declinations. The data at dec. $40^o$ is displayed in Figure 2. We plot the 10 GHz, 15 GHz and 33 GHz scans from top to bottom. In the same figure, the smooth curves correspond to a Wiener filtered data set (Press et al. 1992). Our approach, based on calculating the power spectrum of the original data, allows for the use of filtering techniques to remove noise. Given a prior model and a noise level, the Wiener filter gives an optimal estimate of the background signal. In Figure 2, the Wiener filtered data was constructed using a $n = 1$ prior model and a noise level of $10\mu K$, approximately the noise level quoted by Hancock et al. (1994) for the 33 GHz scan. In Figure 3 we plot the 1D power spectrum obtained from the Fourier transform of the raw data (upper solid curve) and the Wiener filtered data (lower solid curve). The data was binned before computing the power, and the spectrum of the three frequencies was averaged to compare with the results of Table 1. For comparison, in Figure 3 we also plot, with dashed lines, the theoretical power for $m = -2, 1$ ($n = 1, 4$).

One should notice the high frequency tail present in the raw data. This high oscillations are due to the noise present in the original data. They do not have a cosmological origin since above $l = 50$ the window function goes to zero. However, this high frequency tail drives the slope to higher values. To see that this is indeed the case, we estimated the slope for the Wiener filtered data. The effect of the filtering was to fully remove the high frequency tail and to bring the slope close to $m = -2$. In Table 2 we give the slope of the 1D power spectrum obtained from the original Tenerife data for the three different scans of 10, 15 and 33 GHz and from the Wiener filtered data. We should remark that the raw data is compatible with high values of $m$ indicating a non-negligible noise contribution to the temperature anisotropy. For comparison, assuming a prior model of $n = 1$ and a noise level of about 25%, the slope of the power spectrum rises significantly for the scan of 15 GHz and comes close to a value compatible with $m = -2$ ($n = 1$). The average of the 3 scans is, in the raw data, compatible with $m = -2$, even though it centers on values around $m = 0$, indicating that white noise is present in the data. In Table 2 the error bar attached to the slope averaged over three scans comes from the result of the previous Monte Carlo simulations. One can see that from the Tenerife data alone, a cosmological signal of $m \geq 3$ can be ruled out at $1\sigma$ level. On the other hand, for the Wiener filtered data, $m_{1D}$ is very close to $m = -2$. Furthermore, at $1\sigma$ level, the Wiener filtered data is incompatible with $m = 0$. If the Wiener data is an optimal reconstruction of the signal measured by the experiment, this result argues towards a cosmological origin of the temperature anisotropy in the scan.

## 5. Discussion.

In this article we have seen that on small angular experiments information can be obtained about the slope of temperature anisotropy power spectrum. It is clear, from the results of our previous section, that our estimates are less conclusive than COBE/DMR. However, this is not due to analyzing a 1D scan instead of a full 2D map but it is due to the lack of a large number of independent data points in the Tenerife data, about 20 per frequency channel. Our estimation of the radiation anisotropy power spectrum was done independent of other experiments and in a different region of $l$-space than COBE/DMR. Our main conclusion is that the slope is $m < 3$ at $1\sigma$ confidence level.

Experiments like Max and MAXIMA (Hanany et al. 1996) or the combined observations of Tenerife and Bartol–IAC, that sample the same region of the sky but with different resolution, could benefit from an analysis like the one done here. Improvement in the quality of the data and larger data sets would decrease the uncertainties and to improve our estimation of the spectral index. Moreover, the temperature anisotropy power spectrum changes slope very quickly close to the Doppler peaks. For a CDM model with $\Omega_b h^2 = 0.02$ the slope rises from $m = -2$ on large angular scales up to $m = -1$ before decreasing with $m = -3$ after the first Doppler peak. Experiments carried out on the same region of the sky, with high signal to noise ratio, at different frequencies, can be used to probe the existence of a Doppler peak.

**Acknowledgments.** This research was supported by Spanish German Integrated Actions HA 1995 - 0079. FAB would like to acknowledge the support of the Junta de Castilla y León, grant SA94-82. We thank the Tenerife group for kindly providing us with their data.

**REFERENCES**

Bennet, C. L., Banday, A. J., Gorski, K. M. et al. 1996, ApJ, 464, L1

Bond, J. R. 1997 in Evolution of the Universe, ed. G. Börner, S. Gottlöber, to be published (Wiley: New York), preprint astro-ph/9512142

Bond, J. R., & Efstathiou, G. 1987, MNRAS226, 655

Broadhurst, T. J., Ellis, R. S., Koo, D. C., & Szalay, A. S. 1990, Nature, 343, 726

Bunn, E. F., Hoffman, Y., & Silk, J. 1996, ApJ, 464, 1

Ganga, K., Ratra, B., & Sugiyama, N. 1996, ApJ461, L61

Gorski, K., Ratra, B., Stompor, R., Sugiyama, N. & Banday, A. J. 1996 ApJ, submitted, preprint astro-ph/9608054

Gradshteyn, I. S., & Ryzhik, I. M. 1980, Tables of Integrals, Series and Products (Academic Press: San Diego)

Hanany, S. et al. 1996, preprint astro-ph/9609098

Hancock, S. et al. 1994, Nature, 367, 333

Kaiser, N., & Peacock, J. A. 1991, ApJ, 379, 482

Lineweaver, C. H., Barbosa, D., Blanchard, A., & Bartlett, J. G. 1996, Astronomy & Astrophysics, to be published, preprint astro-ph/9610133

Lucy, L. B. 1994, Reviews of Modern Astronomy, 7, 31

Maddox, S. J., Efstathiou, G., Sutherland, W. J. & Loveday, J. 1990, MNRAS, 242, 43P

Peacock, J. A. 1992, in New Insights into the Universe, ed. V. J. Martinez, M. Portilla & D. Sáez (Springer–Verlag: Heidelberg)

Peebles, P. J. E. 1982, ApJ, 263, L1

Piccirillo, L., Femenia, B., Kachwala, N., Rebolo, R., Limon, M., Guttierrez, C. M., Nicholas, J., Schaefer, R. K. & Watson, R. A. 1996, ApJ, to be published, preprint astro-ph/9609186

Press, W. H., Teukolsky, S. A., Vetterling, W. T. & Flannery B. P. 1992, Numerical Recipes (Cambridge University Press: Cambridge)

Sachs, R.K. & Wolfe A.M. 1967, ApJ, 147, 73

Scott, D. et al. 1995, Science, 268, 829

White, M., Scott , D. & Silk, J. 1994, ARA&A, 32, 319

White, M., Viana, P. T. P., Liddle, A. R., & Scott, D. 1996, MNRAS, submitted, preprint astro-ph/9605057

---

| scan | data | Wiener filtered data |
|---|---|---|
| 10 GHz | - 0.070 | - 0.108 |
| 15 GHz | - 0.101 | - 0.185 |
| 33 GHz | - 0.148 | - 0.170 |
| 10+15+33 | $-0.11 \pm 0.05$ | $-0.15 \pm 0.05$ |

Table 2: Slope of the 1D power spectrum for the three Tenerife scans. Column 2 corresponds to the original data and Column 3 corresponds to the Wiener filtered data where the prior model is $n = 1$ and the assumed noise level was $10\mu K$.

| m | $m_{1D}(theo)$ | $m_{1D}(sim)$ | $\sigma_{noise}=0$ | $\sigma_{noise}=0.2\sigma_s$ | $\sigma_{noise}=0.5\sigma_s$ | $\sigma_{noise}=\sigma_s$ |
|---|---|---|---|---|---|---|
| -3.0 | -0.178 | -0.183 | -0.177 | -0.166 | -0.133 | -0.081 |
| -2.5 | -0.165 | -0.169 | -0.171 | -0.162 | -0.133 | -0.086 |
| -2.0 | -0.152 | -0.156 | -0.160 | -0.156 | -0.130 | -0.089 |
| -1.5 | -0.139 | -0.143 | -0.152 | -0.148 | -0.128 | -0.090 |
| -1.0 | -0.127 | -0.131 | -0.141 | -0.138 | -0.122 | -0.089 |
| 0.0 | -0.104 | -0.104 | -0.119 | -0.117 | -0.106 | -0.082 |
| 1.0 | -0.083 | -0.086 | -0.094 | -0.094 | -0.087 | -0.070 |
| 3.0 | -0.047 | -0.050 | -0.052 | -0.052 | -0.049 | -0.041 |

Table 1: In the second column we give the theoretical value of the slope of the 1D power spectrum $P_{1D}(q)$ for the different radiation anisotropy power spectra. In the third column we give the slope of the single realization of the sky used to generate the scans. Columns 4 to 7 give the mean slope of five hundred simulated Tenerife scans. The values were obtained after averaging the power spectrum of three different scans before computing the slope. The r.m.s. deviation was very similar to all simulations ($\tilde{\sigma} \approx 0.04 - 0.05$).

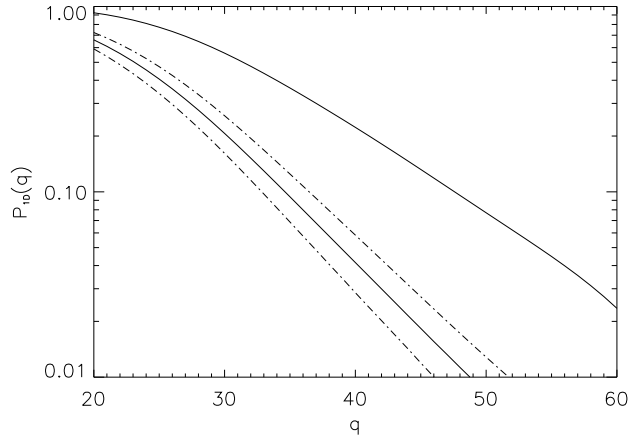

Fig. 1.— One–dimensional power spectra (eq. [6]) as it would be measured from a Tenerife scan. Four different slopes have been considered. The power spectra displayed from left to right corresponds to: $m = -2.5, -2, -1.5, 1$. The most negative $m$ gives rise to the largest negative slope.

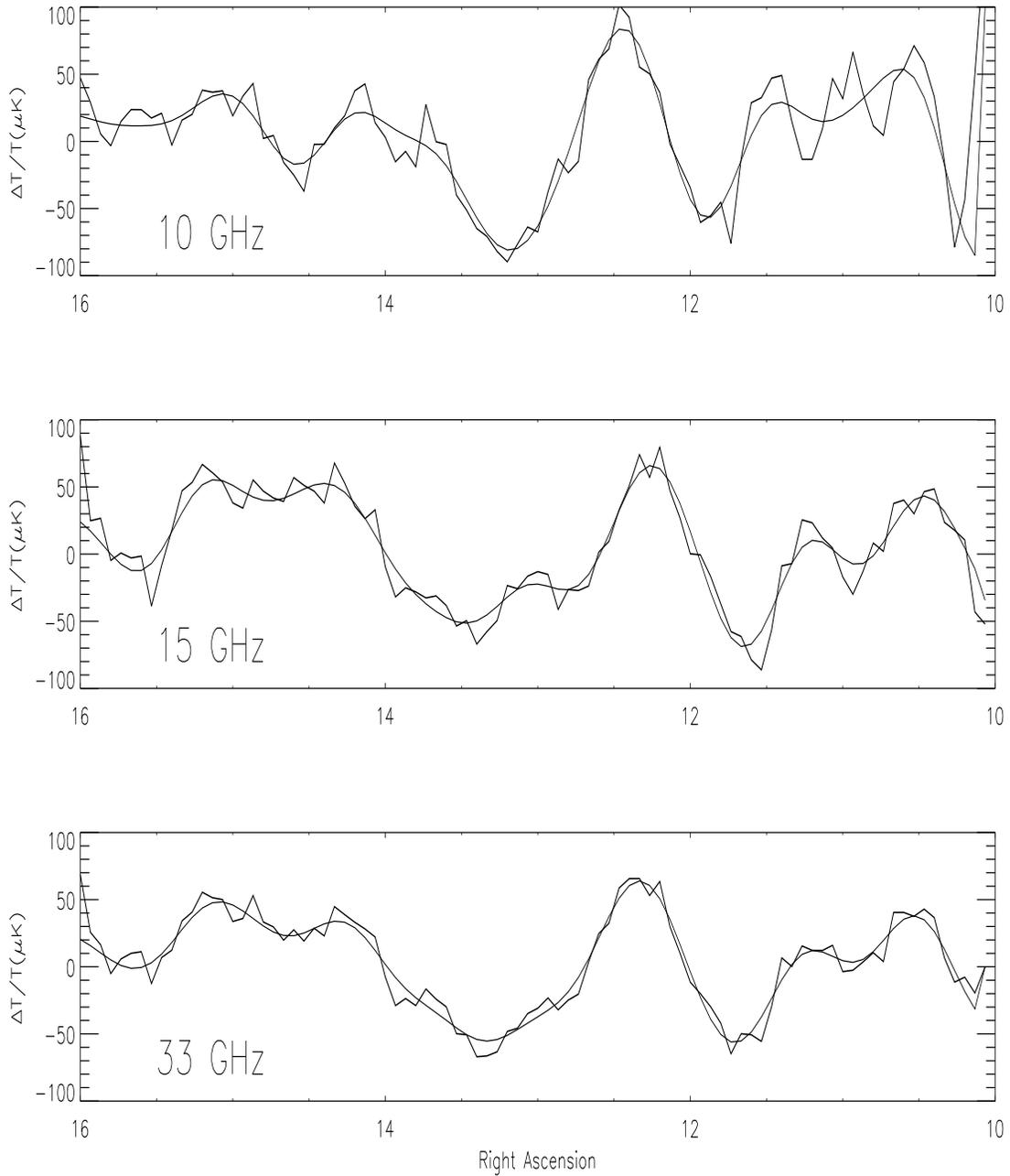

Fig. 2.— Temperature anisotropies of Tenerife scans at 10, 15 and 33 GHz. The smooth curve corresponds to the Wiener filtered data with a $n = 1$ prior model and noise level of $10\mu K$.

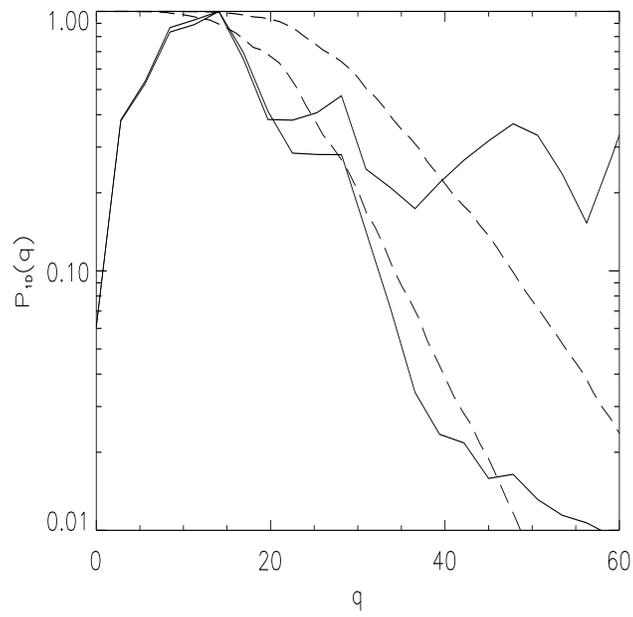

Fig. 3.— The solid lines correspond to the average power spectrum obtained from the three scans of original data (upper solid line) and of Wiener filtered data (lower solid line). Notice the long tail of high frequency noise present in the original data. For comparison, the theoretical one-dimensional power for $n = 1, 4$ (left, right dashed curve) are also plotted.